\documentclass[a4paper]{article}
\pdfoutput=1
\usepackage[english]{babel}
\usepackage[utf8]{inputenc}
\usepackage[T1]{fontenc}
\usepackage{authblk}
\usepackage{siunitx}
\usepackage{amsmath}
\usepackage{graphicx}
\usepackage[colorinlistoftodos]{todonotes}
\usepackage[colorlinks=true, allcolors=black,pdftex]{hyperref}

\title{Absolute spectral metrology of XFEL pulses using diffraction in crystals}
\author[1]{Ilia Petrov}
\author[1]{Liubov Samoylova}
\author[1]{Sarlota Birnsteinova}
\author[1]{Valerio Bellucci}
\author[1]{Mikako Makita}
\author[1]{Tokushi Sato}
\author[1]{Romain Letrun}
\author[1]{Jayanath Koliyadu}
\author[1]{Raphael de Wijn}
\author[2]{Andrea Mazzolari}
\author[2]{Marco Romagnoni}
\author[1]{Richard Bean}
\author[1]{Adrian Mancuso}
\author[3]{Alke Meents}
\author[3,4,5]{Henry N. Chapman}
\author[1,3]{Patrik Vagovic}

\affil[1]{European X-Ray Free-Electron Laser Facility, Holzkoppel 4, D-22869 Schenefeld, Germany}
\affil[2]{INFN Ferrara Division, via Saragat 1, I-44122 Ferrara, Italy}
\affil[3]{Center for Free-Electron Laser Science CFEL, Deutsches Elektronen-Synchrotron DESY, Notkestr. 85, 22607 Hamburg, Germany}
\affil[4]{The Hamburg Centre for Ultrafast Imaging, Luruper Chaussee 149, 22761 Hamburg, Germany}
\affil[5]{Department of Physics, Universität Hamburg, Luruper Chaussee 149, 22761 Hamburg, Germany}

\begin{document}
\maketitle

\begin{abstract}
At modern X-ray sources, such as synchrotrons and X-ray Free-Electron Lasers (XFELs), it is important to measure the absolute value of the photon energy directly. Here, a method for absolute spectral metrology is presented. A photon energy estimation method based on the spectral measurements and rocking of diffracting crystals is presented. The photon energy of SASE1 channel of the European XFEL was measured, and the benefits and applications of the precise photon energy evaluation are discussed.
\end{abstract}

\section{Introduction}
At X-ray Free-Electron Lasers (XFELs), the photon energy is determined by various parameters such as undulator period, magnetic field, electron energy etc. \cite{saldin:book}. However, due to the complexity of measuring these parameters, it is difficult to estimate the photon energy with the precision that is required for experiments. That is, due to a low $\sim$20~eV frequency bandwidth at XFELs \cite{kujala20,katayama_sacla_spec,karvinen12,boesenberg17}, the photon energy needs to be determined with the precision of several eVs to set up the experiment to operate within the highest spectral intensity of XFELs. In particular, the photon energy helps align the crystal optics to the diffraction orientation. In X-ray absorption spectroscopy experiments, the photon energy can be adjusted such that absorption edges are within the highest intensity of X-rays.

Here, we present a direct photon energy measurement method using diffraction in monocrystallines and measurements of spectrum. When inserted into XFEL beam, flat crystals diffract a fraction of the spectrum, such that a dip in the spectrum appears. The angular positions of the crystals during scans allow to determine the resolution of the spectrometer, and, by matching the spectral dip position in opposite diffraction directions, the absolute photon energy. 

\section{Experimental}

\begin{figure}[ht!]
\centering\includegraphics[width=0.7\textwidth]{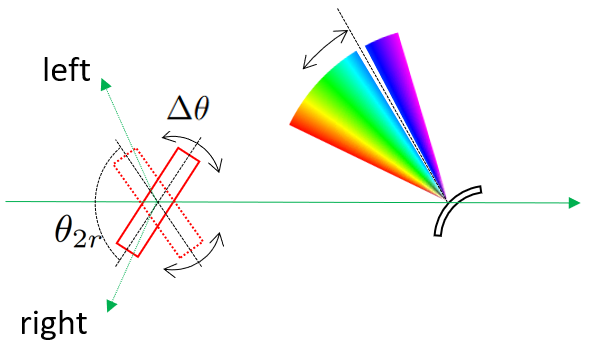}
\caption{The layout of the experiment. The rocking of the crystal (red rectangle) in different orientations relative to the incident beam will provide a dip in the spectrum which will be shifting along photon energies. We denote the opposite diffraction directions as "left" and "right", since the diffraction plane in the experiment was the horizontal plane.}
\label{layout}
\end{figure}

The schematic of the experiment is shown in Fig.~\ref{layout}. A spectrometer based on a strongly bent High-Pressure High-Temperature (HPHT) diamond crystal \cite{boesenberg17} in (440) reflection  was used to measure the Self-Amplified Spontaneous Emission (SASE) spectrum of SASE1 channel at SPB/SFX instrument of European XFEL. When a flat crystal is inserted into the beam upstream of the spectrometer, the photons within the Darwin width of the flat crystal are diffracted which leads to a dip in the spectrum. By matching the location of the dip in two opposite orientations of the flat crystal one can determine the wavelength $\lambda_0$ using Bragg's equation

\begin{equation}
    2d\sin(\theta_{2r}/2)=\lambda_0,
    \label{Bragg_abs}
\end{equation}
where $d$ is the lattice spacing of the reflection, $\theta_{2r}$ is the angle between the orientations of the flat crystal for the same position of the dip in the spectrum. Moreover, by measuring the change of the angle $\Delta\theta$ during scans, we can attribute the shift of the spectral dip on the detector to the photon energy difference $\Delta E$ which can be derived using the differential Bragg equation

\begin{equation}
    \label{diffBragg}
    \Delta E=E_0\cdot\Delta\theta\tan\theta_B,
\end{equation}
where $\theta_B=\theta_{2r}/2$ is the Bragg's angle when $\Delta\theta=0$ and $E_0$ is the photon energy for $\theta_B$. The shift of the dip allows to estimate the pixel energy resolution of the detector used to measure the spectrum. As a result, the photon energy and the pixel resolution provide the absolute calibration of the spectrometer.

\section{Measurements}

\begin{figure}[hbt!]
\centering\includegraphics[width=0.7\textwidth]{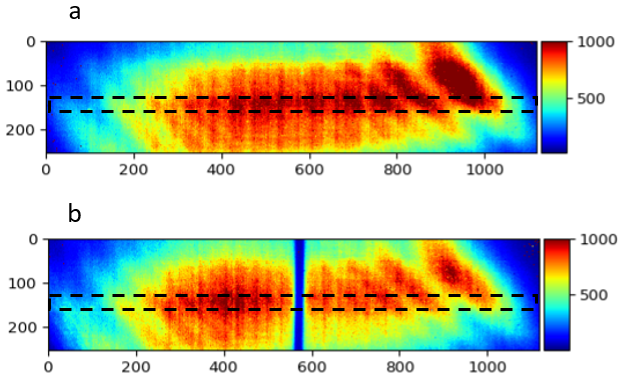}
\caption{a: the image of the spectrum after diffraction from a bent C*(440) crystal that represents the SASE spectrum, b: the image of the spectrum with a flat C*(111) 100~\SI{100}{\micro\meter} inserted before the spectrometer. The spectrum is calculated by integrating the intensity in the black dashed rectangle along the vertical axis. The SASE spectrum was calculated from an average of 100 images. For each scan position, an average of 50 images was used.}
\label{zyla_images}
\end{figure}

Zyla 5.5 sCMOS camera was used to record the spectrum from the strongly bent diamond crystal. Fig.~\ref{zyla_images}a shows the image of the SASE spectrum without a dip. When a flat C*(111) crystal is inserted, the X-rays within a Darwin width are diffracted, and the dip in the spectrum appears in Fig.~\ref{zyla_images}b. The spectral images shown in Fig.~\ref{zyla_images}a are acquired by averaging 100 images, in Fig.~\ref{zyla_images}b - 50 images. Each of the images is an average over a train of 40 pulses that arrived with 1.1 Mhz repetition rate. The inclined features in Fig.~\ref{zyla_images} are attributed to the phase-space configuration of the photon pulse during the measurements and/or the distortions the wavefront by optical elements, which are not expected to have affected the absolute spectral calibration because of a well-pronounced dip in the spectrum.

\begin{figure}[ht!]
\centering\includegraphics[width=\textwidth]{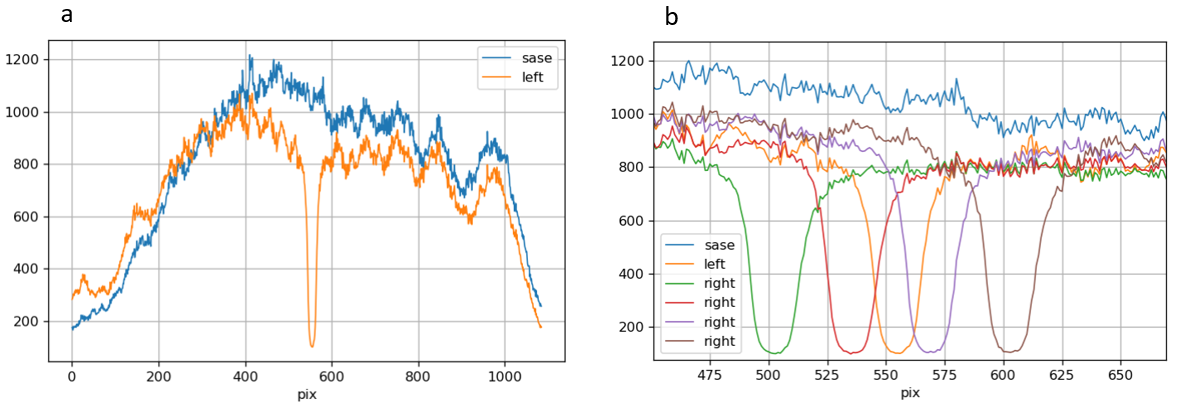}
\caption{a: the measured SASE spectrum (blue) and a spectrum with a dip due to C*(111) diffraction (orange) in a scan with diffraction to the “left” side. b: the measured SASE spectrum (blue) and a spectrum with a dip due to C*(111) diffraction (orange) in a scan with diffraction to the “left” side and dips for various angles in a scan with diffraction to the “right” side, see the legend.}
\label{dips}
\end{figure}

If we integrate over an area of the detector with the strongest signal, as shown in Fig.~\ref{zyla_images}, we will have the spectra with and without the dip due to diffraction shown in Fig.~\ref{dips}a. SmarAct SR-12012 rotation stage was used for the rotation of the flat crystal. The scans were performed in "left" and "right" diffraction orientations since the diffraction was in the horizontal plane. The angle position of the crystal in the "left" diffraction was chosen to be roughly in the center of the spectrum. By selecting the scan angles in the "right" diffraction that provide the dips in the spectrum around the dip in the "left" orientation, as shown in Fig.~\ref{dips}b, using linear interpolation we can estimate the exact angle in the "right" orientation that would correspond to the selected dip in the "left" orientation.

\begin{figure}[ht!]
\centering\includegraphics[width=\textwidth]{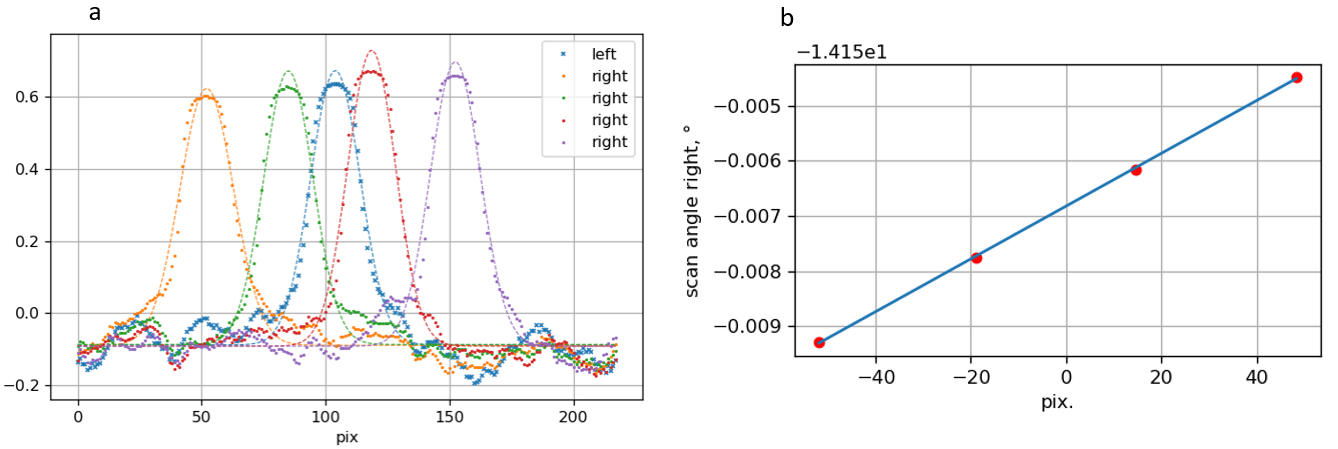}
\caption{a: the restored diffraction peaks in different orientations (dotted lines) and their Gaussian fits (dashed lines). b: the difference between the center of the Gaussian fit of the reflection peak in “left” orientation and in the four “right” orientations in Fig. 5 in pixels (red dots) and their linear fit (blue line).}
\label{peaks_fit}
\end{figure}

In order to restore the position of diffraction peaks in the spectra, we calculate the ratio of the spectra with a dip and the clear SASE spectrum in Fig.~\ref{dips}b, subtract the minimum value of the calculated ratio and multiply it by $-1$. The resulting curves and their Gaussian fits are shown in Fig.~\ref{peaks_fit}a. Linear extrapolation of center positions of Gaussian fits in "right" orientations allows to estimate the scan angle which corresponds to the center of the Gaussian fit of the peak in "left" orientation, as shown in Fig.~\ref{peaks_fit}b. 

The estimated difference of the angles in different orientations that provide the dip in the spectrum at the same photon energy was estimated to be $\theta_{2r}=29.217^{\circ}$ which provides the photon energy $E_0=11935$~eV. The slope of the linear fit in Fig.~\ref{peaks_fit}b allows to estimate the pixel energy resolution to be around $0.038$~eV.  The rotation step of around $0.001^{\circ}$ provided around 1~eV shift of the dip in the spectrum, see Fig.~\ref{peaks_fit}b, and we expect such accuracy of photon energy measurement to be sufficient in view of the SASE bandwidth that amounts to several tens of eV. We also consider the photon energy scaling precision of $0.001$~eV per pixel to be sufficient since such an error would provide around 1~eV difference of the measured SASE bandwidth, which can be tolerated in view of the width of the spectrum.

\section{Conclusions and outlook}
A method for the absolute spectral metrology of XFEL radiation using strongly bent crystals and flat ideal crystals was presented and implemented to precisely measure the photon energy of SASE1 channel of European XFEL. During the experiment, the central photon energy was estimated to be 11935~eV and the estimated bandwidth was around 30~eV.  In practice, since the bandwidth of XFEL radiation is several tens of eV, a 1~eV accuracy of photon energy measurement would be sufficient for typical experiments at XFELs. Angular scans provided the detector photon energy scaling of $0.038$~eV per pixel, and we consider the accuracy of $0.001$~eV to be sufficient in view of the 1~eV difference of SASE bandwidth that such an error would result in. 

\section*{Acknowledgements}
We acknowledge SFX User Consortium for providing Zyla 5.5 sCMOS camera for recording spectral information. The authors acknowledge funding support from Bundesministerium für Bildung und Forschung (BMBF) (05K18XXA), Vetenskapsrådet (VR) (2017-06719), Röntgen Ångström Cluster INVISION project and the funding from  HORIZON-EIC-2021-PATHFINDEROPEN-01-01, Grant agreement: 101046448, MHz-Tomoscopy project.

\bibliographystyle{ieeetr}
\bibliography{encalib}

\newcommand{\noop}[1]{}
\begin{thebibliography}{1}

\bibitem{saldin:book}
E.~L. Saldin, E.~A. Schneidmiller, and M.~V. Yurkov, {\em {The Physics of Free
  Electron Lasers}}.
\newblock Berlin: Springer, 2000.

\bibitem{kujala20}
N.~Kujala, W.~Freund, J.~Liu, A.~Koch, T.~Falk, M.~Planas, F.~Dietrich,
  J.~Laksman, {Maltezopoulos, Th.}, J.~Risch, F.~{Dall'Antonia}, and
  J.~Gr\"unert, ``{Hard x-ray single-shot spectrometer at the European X-ray
  Free-Electron Laser},'' {\em Review of Scientific Instruments}, vol.~91,
  p.~103101, 2020.

\bibitem{katayama_sacla_spec}
T.~Katayama, S.~Owada, T.~Togashi, K.~Ogawa, P.~Karvinen, I.~Vartiainen,
  A.~Eronen, C.~David, T.~Sato, K.~Nakajima, Y.~Joti, H.~Yumoto, H.~Ohashi, and
  M.~Yabashi, ``A beam branching method for timing and spectral
  characterization of hard x-ray free-electron lasers,'' {\em Structural
  Dynamics}, vol.~3, no.~3, p.~034301, 2016.

\bibitem{karvinen12}
P.~Karvinen, S.~Rutishauser, A.~Mozzanica, D.~Greiffenberg, P.~N. Jurani\'c,
  A.~Menzel, A.~Lutman, J.~Krzywinski, D.~M. Fritz, H.~T. Lemke, M.~Cammarata,
  and C.~David, ``{Single-shot analysis of hard x-ray laser radiation using a
  noninvasive grating spectrometer},'' {\em Optics Letters}, vol.~37,
  pp.~5073--5075, 2012.

\bibitem{boesenberg17}
U.~Boesenberg, L.~Samoylova, T.~Roth, D.~Zhu, S.~Terentyev, M.~Vannoni,
  Y.~Feng, T.~B. {van Driel}, S.~Song, V.~Blank, H.~Sinn, A.~Robert, and
  A.~Madsen, ``{X-ray spectrometer based on a bent diamond crystal for high
  repetition rate free-electron laser applications},'' {\em Optics Express},
  vol.~25, pp.~2852--2861, 2017.

\end{thebibliography}

\end{document}